\documentclass[aps,prb,10pt,twocolumn]{revtex4-1}
\usepackage{bm}
\usepackage{amsmath}
\usepackage{amsfonts}
\usepackage{amssymb}
\usepackage{float}
\usepackage[final]{graphicx}
\usepackage[caption = false]{subfig}
\usepackage{dsfont}
\usepackage[colorlinks=true,linktocpage=true,breaklinks=true,citecolor=blue]{hyperref}

\begin{document}
\title{Multipolar orbital relaxation of the $t_{2g}$ states}
\author{Aur\'elien Manchon$^{1}$}
\email{aurelien.manchon@univ-amu.fr}
\author{Xiaobai Ning$^{1,2,3}$}
\author{Chi Sun$^{1}$}
\author{Tetsuya Sato$^4$}
\author{Takeo Kato$^{4}$}
\author{Tatiana G. Rappoport$^{5,6}$}
\affiliation{$^1$Aix-Marseille Univ, CNRS, CINaM, Marseille, France\\$^2$ National Key Laboratory of Spintronics, Hangzhou International Innovation Institute, Beihang University, Hangzhou, China\\$^3$
Fert Beijing Institute, School of Integrated Circuit Science and Engineering, National Key Laboratory of Spintronics, Beihang University, Beijing, China\\
$^4$ Institute for Solid State Physics, University of Tokyo, Kashiwa, 277-8581, Japan\\
$^5$International Iberian Nanotechnology Laboratory (INL), Av. Mestre José Veiga, 4715-330 Braga, Portugal\\
$^6$Centro Brasileiro de Pesquisas Físicas (CBPF), Rua Dr Xavier Sigaud 150, Urca, 22290-180, Rio de Janeiro-RJ, Brazil}

\begin{abstract}
Using a nonperturbative approach, the relaxation rate of orbital dipolar and quadrupolar moments is computed analytically for the $t_{2g}$ states. In the presence of short-range impurities and in the absence of spin-orbit coupling, the orbital relaxation emerges from the competition between momentum scattering and the effect of the crystal field. In the case of weak disorder, the orbital relaxation time is proportional to the momentum scattering time: each scattering event contributes to destroying the orbital moment. In the case of strong disorder, the effect of the crystal field is averaged out, and the orbital relaxation time is inversely proportional to the momentum scattering. We finally find that the dipolar and quadrupolar orbital moments are coupled by the crystal field, resulting in a complex dynamical behavior upon orbital injection.
\end{abstract}
\maketitle

{\em Introduction - } Orbital currents have drawn substantial interest lately as they can be efficiently generated electrically in metals free from heavy elements via the orbital Hall \cite{Go2018,Jo2018,Canonico2020,Bhowal2021,Salemi2022,Pezo2022,Veneri2024} and orbital Rashba-Edelstein effects \cite{Go2017,Manchon2020,Salemi2021,Adamantopoulos2024}. When spin-orbit coupling is present, orbital currents have been shown to drive current-driven torques \cite{Ding2020,Lee2021a,Lee2021b,Yang2024a,Ding2024,Gupta2025}, magnetoresistance effects \cite{Ding2022}, and orbital pumping \cite{Santos2024,Hayashi2024,Wang2025,Huang2025,Sun2025}. The accumulation of orbital moments has been detected using Hanle effects \cite{Sala2023,Aguilar-Pujol2025}, or the optical (orbital) Kerr effect \cite{Lyalin2023,Choi2023}. On the theory side, substantial effort has been made to describe orbital-charge conversion mechanisms \cite{Jo2018,Salemi2022,Go2017,Manchon2020,Salemi2021}, and recent theories have attempted to apprehend the complex dynamics of orbital currents in crystals \cite{Go2020b,Go2021,Han2022,Go2023a}. In spite of these efforts, the compelling experimental evidence of (relatively) long-range orbital currents \cite{Hayashi2023,Bose2023,Moriya2024,Ledesma-Martin2025,Gao2025} remains a puzzling observation, as the orbital relaxation time has been measured around 2 ps \cite{Sala2023,Aguilar-Pujol2025}. Indeed, contrary to the spin angular momentum, which is conserved in the absence of spin-orbit coupling, the orbital moment interacts strongly with the crystal field potential. A recent theory suggests that this interaction leads to a dramatically reduced orbital conductivity compared to charge and spin \cite{Ning2025}. A natural question, therefore, is how the unique interaction between the orbital moment and the crystal field affects the orbital relaxation time.

\begin{figure}[h!]
\includegraphics[width=6cm]{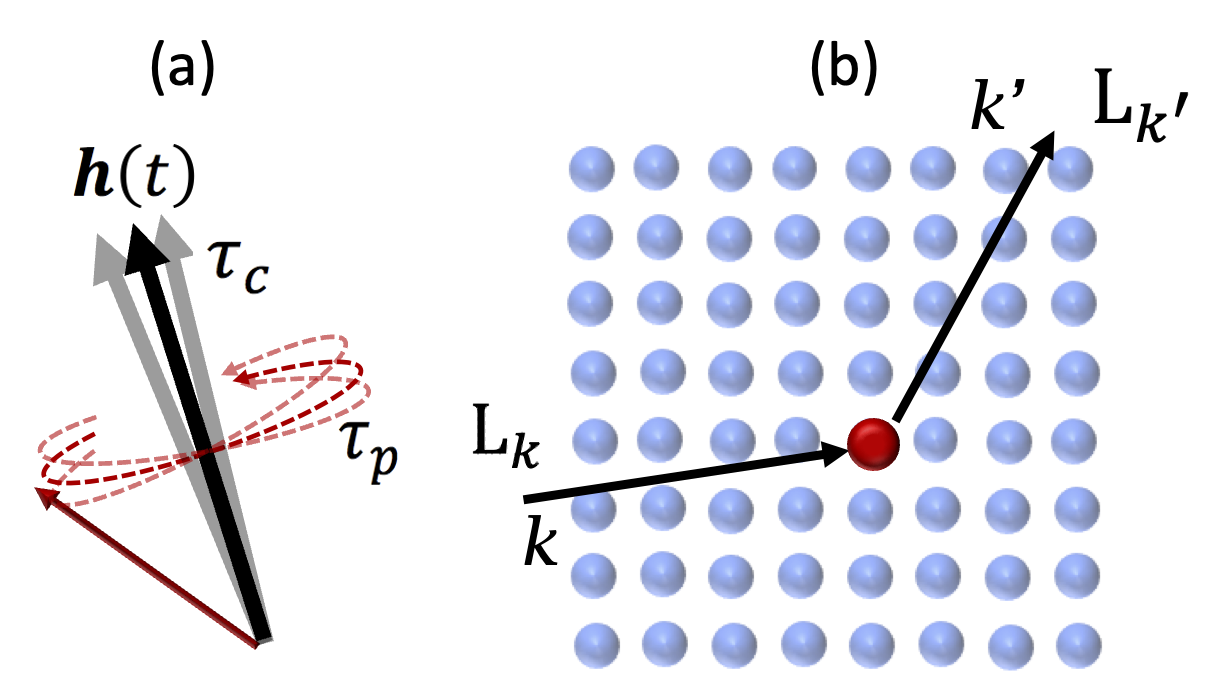}
\caption{(Color online) (a) Spin or orbital relaxation in the presence of a fluctuating magnetic field. $\tau_c$ and $\tau_p$ are the field correlation and spin precession times, respectively. (b) Scattering of a Bloch state with momentum ${\bf k}$ and an orbital moment ${\bf L}_{\bf k}$ towards a state ${\bf k}'$ with an orbital moment ${\bf L}_{\bf k'}$.\label{fig:fig0}}
\end{figure}

The study of spin relaxation in metals and semiconductors has been central to the development of spin-based devices, and the object of comprehensive reviews  \cite{Fabian2007,Dyakonov2008a}. Before addressing the nature of orbital relaxation, it is useful to remind the salient features of spin relaxation.  Following the physical picture developed by D'yakonov \cite{Dyakonov2008a}, the spin relaxation can be understood as arising from the interplay between a fluctuating field ${\bm h}(t)$, characterized by its correlation time $\langle{\bm h}(t){\bm h}(t')\rangle=\tau_c\delta(t-t')$, and the spin precession time $\tau_p$, as depicted in Fig. \ref{fig:fig0}(a). In real materials, this fluctuating field typically represents the spin-orbit field, random magnetic impurities, the nuclear spin, etc. If $\tau_p\gg\tau_c$, the spin remains aligned on the magnetic field and therefore, the spin relaxation time is simply given by the correlation time, $\tau_s\sim \tau_c$. This corresponds to the Elliott–Yafet mechanism\cite{Elliott1954,Yafet1963}, where impurity or phonon scattering can flip the spin, so that increased momentum scattering enhances spin relaxation. If $\tau_c\gg\tau_p$, the effect of the fluctuations is averaged out over one precession and the overall relaxation time is $\tau_s\sim \tau_p^2/\tau_c$. In materials where spin-momentum locking is present (e.g., due to Rashba spin-orbit coupling), the momentum scattering averages out the effective spin-orbit field, leading to a spin relaxation time proportional to the {\em inverse} of the momentum scattering time. This relaxation mechanism is referred to as the D'yakonov-Perel' mechanism \cite{Dyakonov1972}. These effects are generally accounted for in a unified manner using perturbation theory \cite{Boross2013,Szolnoki2017}.

 In the case of orbital transport, since the Bloch state is constructed from the superposition of atomic orbitals, the orbital moment is naturally momentum-dependent. Consequently, each scattering event ${\bf k}\rightarrow{\bf k}'$ is accompanied by a change in the orbital moment ${\bf L}_{\bf k}\rightarrow{\bf L}_{\bf k'}$, see Fig. \ref{fig:fig0}. Hence, one can speculate that momentum scattering might, if strong enough, average out the effect of the crystal field and preserve the orbital moment, as suggested by Ref. \onlinecite{Sohn2024}.

In this Letter, we develop a nonperturbative theoretical method to analytically describe the relaxation of orbital moments of the $t_{2g}$ states. In the absence of spin-orbit coupling, the orbital relaxation arises from the competition between momentum scattering and the crystal field. For weak disorder, it follows the Elliott-Yafet scaling, being proportional to the momentum scattering time. For strong disorder, the effect of the crystal field is averaged out, and the relaxation follows the D’yakonov–Perel' scaling, becoming inversely proportional to the scattering time. This nonlinear dependence implies unconventional temperature dependence. Remarkably, we show that lowering the crystal field symmetry leads to a direct coupling between orbital dipolar and quadrupolar moments, producing an oscillatory dynamics that is expected to impact orbital transport.



{\em Theory of orbital relaxation - } Because the spin angular momentum is a good quantum number in nonmagnetic metals free from spin-orbit coupling, spin relaxation is usually studied using perturbation theory \cite{Fabian2007,Dyakonov2008a,Boross2013,Szolnoki2017}. This approach cannot be applied to the orbital counterpart because it is not an eigenstate of the Hamiltonian. To circumvent this issue, we solve the orbital relaxation analytically using the quantum kinetic approach \cite{Rammer1986}. We consider a cubic crystal in the spherical approximation, i.e., valid to describe $t_{2g}$ states close to the $\Gamma$ point, 
\begin{eqnarray}\label{eq:H}
\hat{\cal H}_k= tk^2\hat{I}+r({\bf k}\cdot\hat{\bf L})^2.
\end{eqnarray}
The first term is the kinetic energy ($t=\hbar^2/2m^*$, with $m^*$ being the effective mass), the second term is the crystal field close to the $\Gamma$ point. Notice that $\hat{\bf L}$ is the operator of orbital angular moment $l=1$ and $\hat{I}_3$ is the unit 3$\times$3 matrix. In the remainder of the manuscript, $\hat{}$ indicates a 3$\times$3 matrix. This Hamiltonian has been widely used to investigate orbital transport \cite{Bernevig2005b,Han2022,Sohn2024} and its eigenstates are characterized by their helicity, $\lambda=(\hbar/k){\bf k}\cdot{\bf L}$. 

We start our analysis from the quantum kinetic equation derived from Keldysh-Dyson equations \cite{Rammer1986,Wang2014},
\begin{eqnarray}
i\hbar\partial_t \hat G^<_k+[\hat G^<_k,\hat {\cal H}_k]=\hat \Sigma^<\hat G^A_k-\hat G^<_k\hat \Sigma^A-\hat G^R_k\hat \Sigma^<+\hat \Sigma^R\hat G^<_k,
\end{eqnarray}
where $\hat G^{<,R,A}_k$ are the lesser, retarded, and advanced Green's functions. We consider short-range, delta-like impurities, $\hat V_{imp}=\sum_jV_0\hat{I}\delta({\bf r}-{\bf R}_j)$, so the self-energies are defined as $\hat \Sigma^{<,R,A}=n_iV_0^2\int\frac{d^3{\bf k}}{(2\pi)^3}\hat G^{<,R,A}_k$, with $n_i$ being the density of impurities. We first perform an energy integration, defining the density matrix as $\hat g^<_k=(1/2i\pi)\int d\varepsilon \hat G^<_k$ and $\hat{\rho}=\int\frac{d^3{\bf k}}{(2\pi)^3}\hat g^<_k$. Posing $\hat\Sigma^{R(A)}=\mp i\hat\Sigma$ and integrating over the energy, we obtain,
\begin{eqnarray}\label{eq:qke}
i\hbar\partial_t \hat g^<_k+i\{\hat g^<_k,\hat\Sigma\}+[\hat g^<_k,\hat {\cal H}_k]=n_iV_0^2(\hat \rho\hat G^A_k-\hat G^R_k\hat \rho).
\end{eqnarray}
We intend to solve this equation analytically to obtain the decay rate of the density matrix $\partial_t\hat \rho$. We are particularly interested in describing the dynamics of the orbital moment dipolar components, $\hat L_i$, with $i=x, y,z$, as well as the quadrupolar components,  $\hat Q_{xy}=1/2\{\hat L_x,\hat L_y\}$, $\hat Q_{z^2}=3\hat L_z^2/2-L^2/2$, $\hat Q_{x^2-y^2}=\hat L_x^2-\hat L_y^2$, etc.

{\em Three-dimensional case -} We first consider a three-dimensional crystal. The unperturbed retarded Green's function reads
\begin{eqnarray}
\hat G_k^{R0}&=&\left((\varepsilon+i0^+)\hat{I}-\hat{\cal H}_k\right)^{-1},\nonumber\\
&=&\frac{({\bf k}\cdot\hat{\bf L})^2/k^2}{\varepsilon+i0^+-\varepsilon_k^r}+\frac{\hat{I}-({\bf k}\cdot\hat{\bf L})^2/k^2}{\varepsilon+i0^+-\varepsilon_k^t},
\end{eqnarray} 
where $\varepsilon_k^t=tk^2$, $\varepsilon_k^r=(t+r)k^2$ are the eigenenergies.

In three dimensions, the density of states reads
\begin{eqnarray}
{\cal N}=\frac{1}{\pi}{\rm Tr}\int\frac{d^3{\bf k}}{(2\pi)^3}G_k^{A}=\frac{\sqrt{\varepsilon}}{2\pi^2}\left(\frac{1}{(t+r)^{3/2}}+\frac{1}{2t^{3/2}}\right),
\end{eqnarray} 
and the self-energy is
\begin{eqnarray}
\hat \Sigma&=&\frac{n_iV_0^2}{3}\pi{\cal N}\hat{I}=\Gamma\hat{I}.
\end{eqnarray} 

We then obtain
\begin{eqnarray}
\left(i\hbar\partial_t+2i\Gamma\right) \hat g^<_k+[\hat g^<_k,\hat {\cal H}_k]=\frac{3\Gamma}{\pi{\cal N}}\left(\hat{\rho}\hat G^A_k-\hat G^R_k\hat{\rho}\right),
\end{eqnarray}
Let us now apply the Fourier transform in time, and, defining $\Omega=\omega+2i\Gamma$, we need to solve the equation
\begin{eqnarray}\label{eq:rec}
 \hat g^<_k=\frac{1}{\Omega}[\hat {\cal H}_k,\hat g^<_k]+\frac{1}{\Omega}\hat {\cal K},\;\hat {\cal K}=\frac{3\Gamma}{\pi{\cal N}}\left(\hat{\rho}\hat G^A_k-\hat G^R_k\hat{\rho}\right).
\end{eqnarray}
Applying the commutator recursively, one obtains a formal expression of $ g^<_k$ as a function of $\hat{\rho}$,
\begin{eqnarray}\label{eq:1}
 \hat g^<_k=\frac{\hat {\cal K}}{\Omega}+\sum_{n\geq 1}\left(\frac{1}{\Omega}\right)^n\left[\hat {\cal H}_k,\frac{\hat {\cal K}}{\Omega}\right]^{(n)},
\end{eqnarray}
where the exponent $^{(n)}$ means the commutator has been applied $n$ times. This expression is equivalent to the common "ladder approximation" in the linear response theory \cite{Rammer1986} and converges for "strong enough" disorder broadening $\Gamma$. Performing this algebra can be very cumbersome, but the high symmetry of the spherical approximation substantially simplifies the calculations. We now assume that a nonequilibrium orbital moment density $\hat\rho_X$ is injected in the system and track its relaxation, with $\hat X=\hat L_x, \hat L_y, ...$. Since $x,y,z$ are equivalent directions, it is sufficient to investigate the dynamics of the components of the density matrix along $\hat L_z$, $\hat Q_{z^2}$, and $\hat Q_{xy}$. The detail of the derivation is given in the Appendix. For $\hat L_z$, we obtain  
\begin{eqnarray}\label{eq:2}
\frac{1}{\tau_{L_z}}=\frac{1}{2\tau}\left(\frac{{\cal N}_t}{{\cal N}}\frac{\xi_t^2}{1+\xi_t^2}+\frac{{\cal N}_r}{{\cal N}}\frac{\xi_r^2}{1+\xi_r^2}\right).
\end{eqnarray}
with $\xi_{r,t}=\tau/\tau_{r,t}$, $\tau=\hbar/2\Gamma$, $\tau_t=\hbar r/t\varepsilon$ and $\tau_r=\hbar r/(t+r)\varepsilon$. We also define the density of states for the two states that contribute to the relaxation, ${\cal N}_t=\sqrt{\varepsilon}/(4\pi^2t^{3/2})$ and  ${\cal N}_r=\sqrt{\varepsilon}/(4\pi^2(t+r)^{3/2})$.

\begin{figure}
\includegraphics[width=8cm]{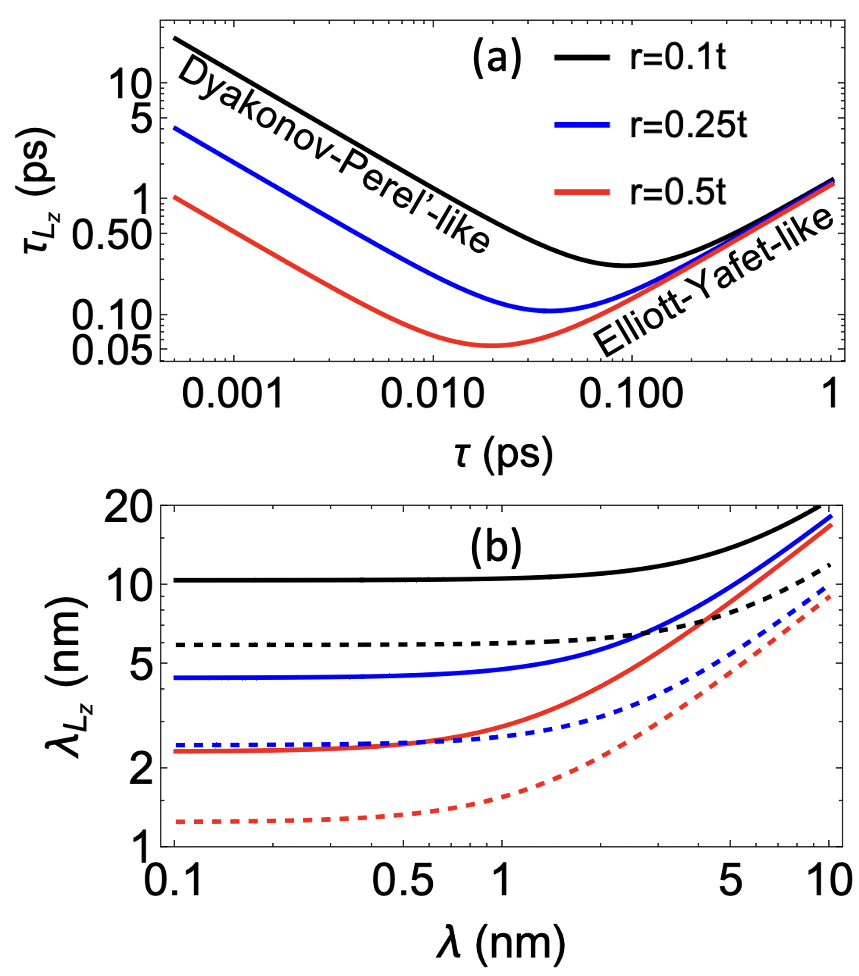}
\caption{(Color online) (a) Orbital relaxation time as a function of the momentum relaxation time for different values of the crystal field $r$. (b) Longitudinal (solid) and transverse (dashed) orbital relaxation length as a function of the mean free path.}\label{fig1}
\end{figure}

 The quadrupole relaxation time is simply
\begin{eqnarray}\label{eq:q2}
\frac{1}{\tau_{Q_{z^2}}}=\frac{1}{\tau_{Q_{xy}}}=\frac{3}{5}\frac{1}{\tau_{L_z}}.
\end{eqnarray}
The relaxation time that we obtain exhibits a nontrivial dependence on the momentum relaxation time $\tau$, suggesting different mechanisms dominate depending on the transport regime, weakly or strongly disordered. The relaxation time for $\hat L_z$ is reported in Fig. \ref{fig1}(a) as a function of $\tau$, for different values of the crystal field $r$. For these calculations, we set $\varepsilon=1.5$ eV and the values of $t$ and $r$ are set to those of Silicon, $t\approx1.5\times10^{-19}$ eV$\cdot$m$^2$ and $r\approx0.5t$ \cite{Bernevig2005b}.

In the strongly disordered regime, we see that the orbital relaxation is first {\em inversely} proportional to $\tau$, and becomes proportional to $\tau$ as soon as $\tau>\tau_{r}$. At long momentum relaxation time, the interaction of the orbital momentum with the crystal field, quantified by $\tau_{r,t}$, is large and relaxes the orbital moment following a process resembles, but is distinct from, Elliott-Yafet: the stronger the scattering, the shorter the lifetime. As soon as $\tau$ is shorter than the interacting time around the crystal field, it averages out this effect, preserving the orbital moment. A similar crossover was reported in real-time spin polarization decay simulations\cite{Culcer2007}. It is interesting to compare our theory with Sohn et al. \cite{Sohn2024}. In their simulation, that extends in the ultrashort timescale $\tau\approx0.5-5$ fs, they observe a sharp increase of the orbital relaxation time with the momentum relaxation time at very short time scale, $\tau_o\sim\tau$ (so, an Elliott-Yafet behavior), and only for $\tau>1.5$ fs, a decrease of the orbital relaxation time with the momentum relaxation time (so, a D'yakonov-Perel' behavior). Our results agree with theirs in the intermediate timescale 10 fs $>\tau>1.5$ fs, but differ for long enough momentum relaxation time as this regime is not covered by Sohn et al. \cite{Sohn2024} simulations. 
 
\begin{figure}
\includegraphics[width=9cm]{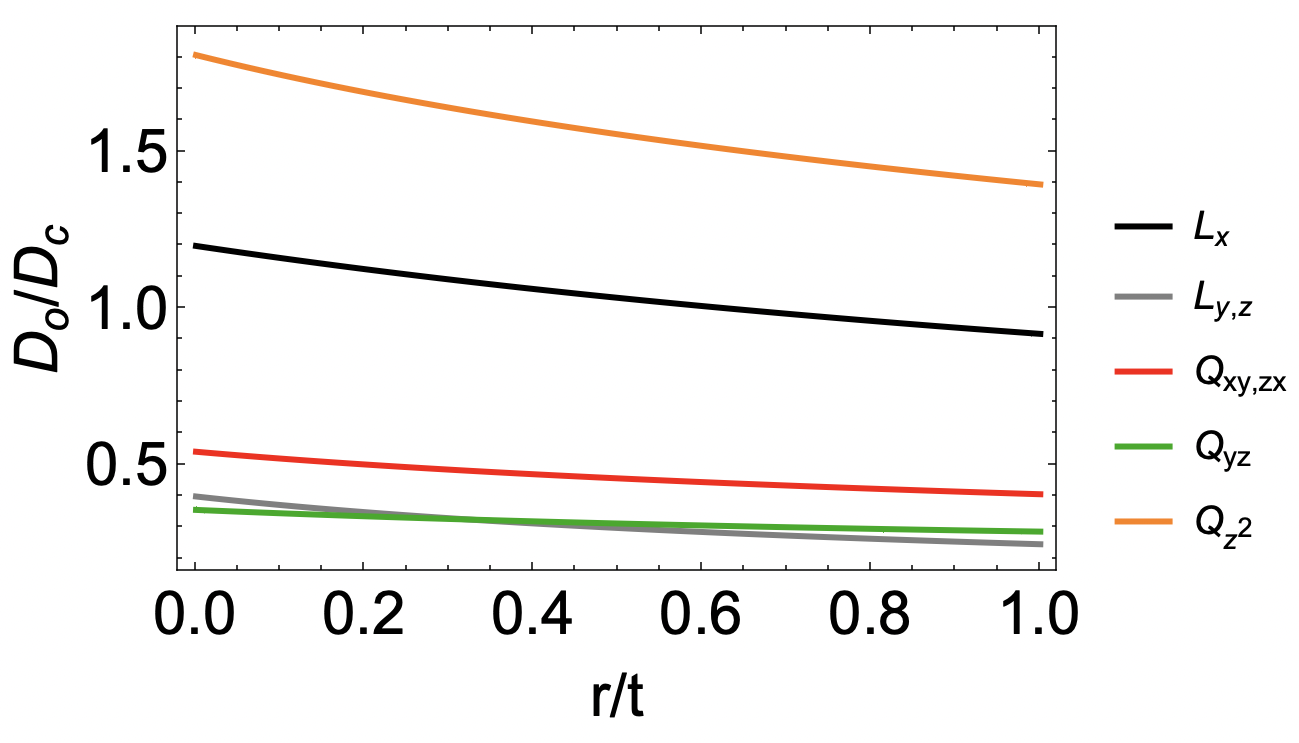}
\caption{(Color online) Diffusion coefficients of the different orbital components, as indicated in the legend, as a function of the crystal field parameter $r/t$. The diffusion coefficients are normalized to the charge diffusivity.}\label{fig2}
\end{figure}

To obtain the orbital relaxation length, we need to compute the diffusion coefficient. To do so, we use the theory developed in Ref. \onlinecite{Ning2025}, and determine the charge diffusivity,
\begin{eqnarray}\label{eq:3}
{\cal D}_{ij}=\frac{2\Gamma}{\pi{\cal N}}\int \frac{d^3{\bf k}}{(2\pi)^3}{\rm Re}\left[v_j{\rm Im}\left(G_0^RL_\alpha G_0^A v_i G_0^A\right)\right],
\end{eqnarray}
and the orbital diffusivity,
\begin{eqnarray}\label{eq:4}
{\cal D}_{i,\alpha}^{j,\beta}=\frac{2\Gamma}{\pi{\cal N}}\int \frac{d^3{\bf k}}{(2\pi)^3}{\rm Re}\left[{\cal J}_{j}^{l,\beta}{\rm Im}\left(G_0^RL_\alpha G_0^A v_i G_0^A\right)\right],
\end{eqnarray}
with ${\cal J}_{j}^{l,\beta}=(1/2)\left\{v_j,L_\beta \right\}$. Explicitly, we obtain the charge diffusivity
\begin{eqnarray}
{\cal D}_c=\frac{2}{\pi^2\hbar{\cal N}\Gamma}\left(\frac{\varepsilon^{3/2}}{\sqrt{t+r}}+\frac{\varepsilon^{3/2}}{2\sqrt{t}}\right),
\end{eqnarray}
and the orbital diffusivities when the orbital moment is aligned with the propagation direction, ${\cal D}_l^\|$, and when it is transverse to it, ${\cal D}_l^\bot$,
\begin{eqnarray}\label{eq:5}
{\cal D}_l^\|=\frac{2}{5\pi^2\hbar{\cal N}\Gamma}(6t+7r)\left(\frac{\varepsilon}{t+r}\right)^{3/2},\\
{\cal D}_l^\bot=\frac{1}{10\pi^2\hbar{\cal N}\Gamma}(4t+3r)\left(\frac{\varepsilon}{t+r}\right)^{3/2}.
\end{eqnarray}
The diffusion coefficients for the orbital quadrupoles can be obtained similarly,
\begin{eqnarray}\label{eq:q5}
{\cal D}_{Q_{xy,zx}}=\frac{1}{70\pi^2\hbar{\cal N}\Gamma}\left(\frac{52t+47r}{2(t+r)^{3/2}}+\frac{12t-r}{t^{3/2}}\right)\varepsilon^{3/2},\\
{\cal D}_{Q_{yz}}=\frac{1}{70\pi^2\hbar{\cal N}\Gamma}\left(\frac{51t+46r}{2(t+r)^{3/2}}+\frac{2t+r}{t^{3/2}}\right)\varepsilon^{3/2},\\
{\cal D}_{Q_{z^2}}=\frac{1}{21\pi^2\hbar{\cal N}\Gamma}\left(\frac{156t+155r}{6(t+r)^{3/2}}+\frac{12t-r}{t^{3/2}}\right)\varepsilon^{3/2},
\end{eqnarray}
and ${\cal D}_{Q_{x^2-y^2}}=4{\cal D}_{Q_{xy,zx}}$. These coefficients, normalized to the charge diffusivity, are displayed in Fig. \ref{fig2} as a function of the crystal field parameter $r$. ${\cal D}_l^\|/{\cal D}_c$ decreases with the magnitude of the crystal field, $r$, but whereas the orbital diffusivity is generally smaller than the charge diffusivity (with the notable exception of ${\cal D}_{Q_{z^2}}$), it is not as small as predicted in Ref. \onlinecite{Ning2025} for vanadium and tantalum. We attribute the relatively "good" orbital diffusivity of the $t_{2g}$ states to the high symmetry of the crystal field. We expect that crystal fields with lower symmetry, such as cubic or hexagonal, shall further reduce the diffusivity of the orbital moment. This question is left to future studies.
The mean free path is defined $\lambda=\sqrt{{\cal D}_c(\hbar/2\Gamma)}$ and the orbital relaxation length is $\lambda_o^{\|,\bot}=\sqrt{{\cal D}_l^{\|,\bot}\tau_o}$. They are computed in Fig. \ref{fig2}. For strong scattering, the orbital relaxation length is mostly independent of the mean free path, whereas it is proportional to it for cleaner systems. In the disorder regime, we obtain an orbital relaxation length that can be an order of magnitude larger than the mean free path. Although this is a qualitatively interesting result, it needs to be taken with sane care given the simplicity of the present model. An interesting feature, though, is the strong anisotropy of the relaxation length when the orbital moment is oriented along (solid lines) or perpendicular (dashed lines) to the propagation direction. This anisotropy in the transport is a hallmark of orbital transport.

{\em Two dimensional case -} Let us move on to the two-dimensional case, setting $k_z=0$ in Eq. \eqref{eq:H}. The density of states and self-energies are then
\begin{eqnarray}
{\cal N}&=&=\frac{1}{2\pi}\left(\frac{1}{t+r}+\frac{1}{2t}\right),\;\hat \Sigma=\Gamma_0\left(\hat{I}+\beta\hat L_z^2\right),\\
\Gamma_0&=&\frac{n_iV_0^2}{4}\frac{1}{r+t},\;\beta=\frac{r}{2t}.
\end{eqnarray} 
The important difference with the 3D situation is that the system's symmetry is lowered from spherical to cylindrical, lifting the degeneracy between in-plane and out-of-plane orbitals. As a result, the self-energy acquires a new term, $\sim\beta$, that needs to be accounted for in our recursive procedure. After some algebra, we find that Eq. \eqref{eq:1} remains valid by setting
 $\Omega=\omega+2i\Gamma_0(1+\beta)$ for the components $\hat L_z$, $\hat Q_{z^2}$, $\hat Q_{xy}$ and $\hat Q_{x^2-y^2}$ and $\Omega=\omega+2i\Gamma_0(1+\beta/2)$ for $\hat L_x$, $\hat L_y$, $\hat Q_{yz}$ and $\hat Q_{zx}$. Therefore, the relaxation time of the out-of-plane component reads
  \begin{eqnarray}\label{eq:6}
\frac{1}{\tau_{L_z}}=\frac{3}{4\tau}\left(\frac{{\cal N}_r}{{\cal N}}\frac{(\xi^\bot_{r})^2}{1+(\xi^\bot_{r})^2}+\frac{{\cal N}_t}{{\cal N}}\frac{(\xi^\bot_t)^2}{1+(\xi^\bot_t)^2}\right),
\end{eqnarray}
with $\xi^\bot_{r}=(\varepsilon/2\Gamma_0(1+\beta))(r/(r+t))$, and $\xi^\bot_t=(\varepsilon/2\Gamma_0(1+\beta))(r/t)$. We define the average broadening $\Gamma=(1/3){\rm Tr}\hat \Sigma=\Gamma_0(1+\frac{2}{3}\beta)=\hbar/2\tau$, as well as the density of states for the two states that contribute to the relaxation, ${\cal N}_t=1/(4\pi t)$ and  ${\cal N}_r=1/(4\pi (t+r))$ and ${\cal N}={\cal N}_t+2{\cal N}_r$. We also find that the relaxation of the quadrupole moments read $\frac{1}{\tau_{Q_{x^2-y^2,xy}}}=\frac{1}{2\tau_{L_z}}$. The relaxation of these components is very similar to the 3D case and does not present any peculiarity. However, this is not the case of the in-plane components.
\begin{figure}
\includegraphics[width=8cm]{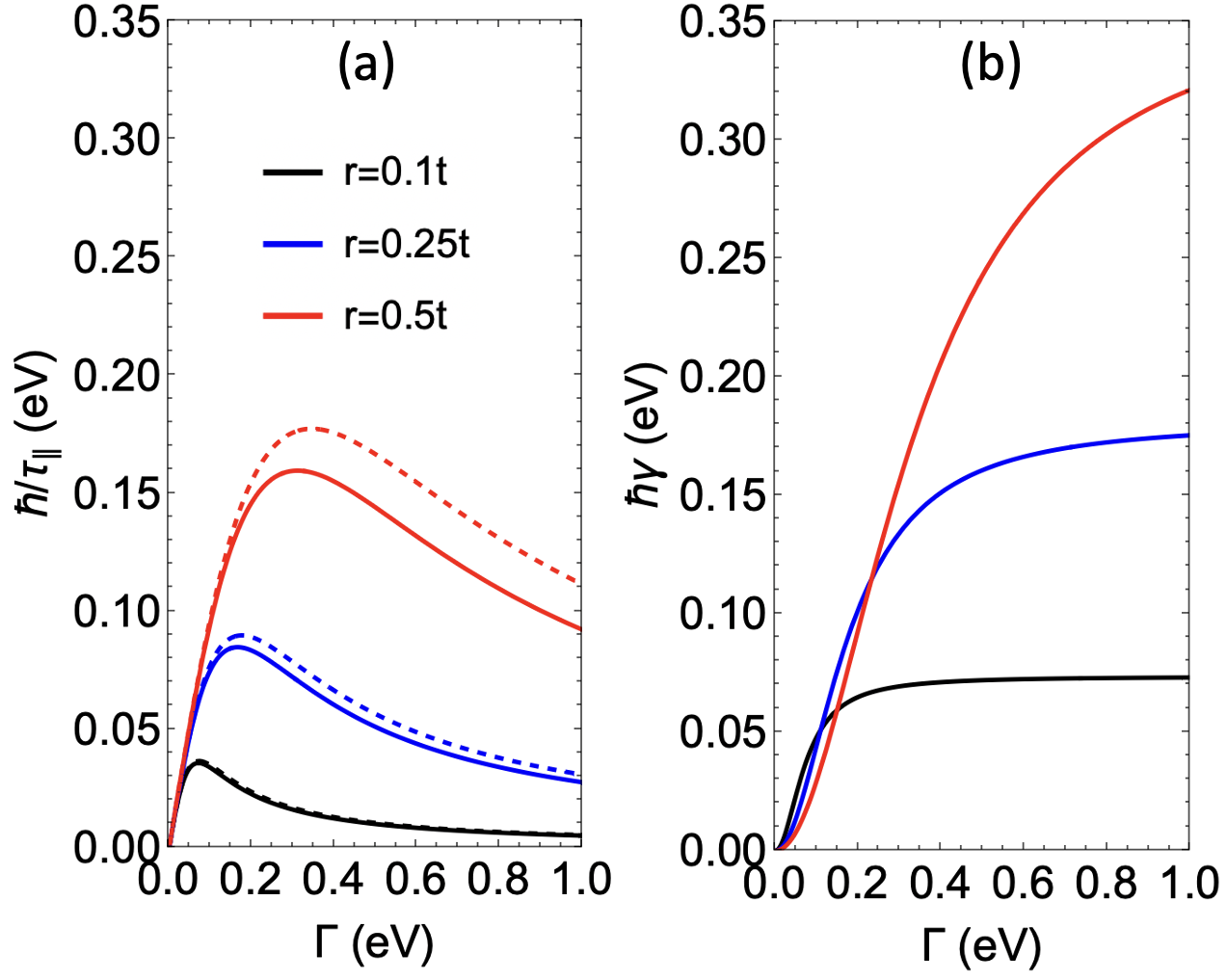}
\caption{(a) Orbital relaxation rate and (b) dipole-quadrupole coupling constant a function of the momentum scattering rate for different values of $r$.\label{fig:fig3}}
\end{figure}
We find that the planar dipole components $L_x,\;L_y$ are coupled to their quadrupole counterpart $Q_{yz},\;Q_{zx}$, respectively,
\begin{eqnarray}\label{eq:7}
\partial_t\rho_{L_x}&=&-\gamma \rho_{Q_{yz}}-\frac{1}{\tau_{\|}}\rho_{L_x},\\
\partial_t\rho_{Q_{yz}}&=&-\frac{1}{\tau_{\|}} \rho_{Q_{yz}}+\gamma\rho_{L_x},
\end{eqnarray}
and 
\begin{eqnarray}\label{eq:8}
\partial_t\rho_{L_y}&=&\gamma \rho_{Q_{zx}}-\frac{1}{\tau_{\|}}\rho_{L_y},\\
\partial_t\rho_{Q_{zx}}&=&-\frac{1}{\tau_{\|}} \rho_{Q_{zx}}-\gamma\rho_{L_y}.
\end{eqnarray}
The coupling constant and the relaxation time are
\begin{eqnarray}\label{eq:9}
\gamma &=&\frac{3}{8\tau}\left(\frac{{\cal N}_r}{{\cal N}}\frac{\xi_r^{\parallel}}{1+\left(\xi_r^{\parallel}\right)^2}+\frac{{\cal N}_t}{{\cal N}}\frac{\xi_t^{\parallel}}{1+\left(\xi_r^{\parallel}\right)^2}\right),\\
\frac{1}{\tau_{\|}}&=&\frac{3}{8\tau}\left(\frac{{\cal N}_r}{{\cal N}}\frac{\left(\xi_r^{\parallel}\right)^2}{1+\left(\xi_r^{\parallel}\right)^2}+\frac{{\cal N}_t}{{\cal N}}\frac{\left(\xi_t^{\parallel}\right)^2}{1+\left(\xi_t^{\parallel}\right)^2}\right),
\end{eqnarray}
where $\xi_r^{\|}=(\varepsilon/2\Gamma_0(1+\beta/2))(r/(r+t))$, and $\xi_t^{\|}=(\varepsilon/2\Gamma_0(1+\beta/2))(r/t)$. The coupling $\gamma$ is responsible for the oscillation of the dipole and quadrupole moments. In other words, when an orbital dipole moment is injected in a crystal, it experiences a damped oscillation around the crystal field, $\rho_{L_x}\sim e^{-t/\tau_{\parallel}}\cos\left(\gamma t+\phi\right)$, accompanied by the creation of a nonequilibrium quadrupole moment, $\rho_{Q_{yz}}\sim e^{-t/\tau_{\parallel}}\sin\left(\gamma t+\phi\right)$. The dependence of the relaxation times for planar and perpendicular orbital moments and that of the coupling constant is represented in Fig. \ref{fig:fig3} as a function of the disorder broadening. The coupling between dipole and quadrupole moments is generally stronger than the relaxation time and increases with the disorder, in sharp contrast with the relaxation time. This dynamics is markedly different from that of a spin subjected to a magnetic field. For a field along $z$, injecting a spin component $S_x$ produces a component $S_y$, inducing therefore a precession. In the case of the orbital dipole, injecting $L_x$ does not generate $L_y$ but rather $Q_{yz}$, reflecting the fact that the orbital moment is not conserved during the propagation. Consequently, not only does the orbital dipole moment oscillate in time and space, but it is also accompanied by the generation of an orbital quadrupole moment of the same magnitude. 

{\em Conclusions -} We have shown analytically that the orbital relaxation rate is driven by a D’yakonov–Perel-type mechanism, and exhibits a crossover from $\sim\tau$ to $\sim1/\tau$. In other words, our theory suggests that disorder can preserve orbital moments, potentially explaining the long relaxation lengths observed experimentally. Remarkably, in this regime, the orbital diffusion length becomes independent of the mean free path, leading to a temperature-independent behavior. Our model, however, considers only short-range impurities and neglects phonon scattering. Furthermore, we find that when the crystal field symmetry is lowered, dipolar and quadrupolar orbital moments interconvert, leading to damped oscillatory dynamics. This finding is particularly intriguing, as it raises the possibility of employing quadrupolar moments as time-reversal-even information carriers.

\begin{acknowledgments}
A.M. thanks Prof. H.W. Lee for insightful discussions. X.N. was supported by the China Scholarship Council Program. A. M., C.S., and T.G.R. were supported by EIC Pathfinder OPEN grant 101129641 “OBELIX”. A.M. and C.S. were supported by the France 2030 government investment plan managed by the French National Research Agency under grant reference PEPR SPIN – [SPINTHEORY] ANR-22-EXSP-0009. T.G.R. thanks the support from CNPq (Grant No 305013/2024-6). T.K. and T.S. were supported by Grants-in-Aid for Scientific Research (Grants No. JP23KJ0702 and No. JP24K06951) and the Japan Science and Technology Agency (JST) ASPIRE Program No. JPMJAP2410.
\end{acknowledgments}

\newpage
\begin{widetext}
\appendix

\newpage
\section{Derivation of the relaxation rate}

Here, we present the details of the derivation of Eq. \ref{eq:2}. Starting from Eq. \eqref{eq:1}, we define 
\begin{eqnarray}
\hat {\cal K}_X=\frac{3\Gamma}{\pi{\cal N}k^2}\left({\cal A}\{\hat X,({\bf k}\cdot\hat{\bf L})^2\}+{\cal B}[\hat X,({\bf k}\cdot\hat{\bf L})^2]+k^2{\cal C}\hat X\right),
\end{eqnarray}
with
\begin{eqnarray}
&&{\cal A}=\frac{i\Gamma}{(\varepsilon-(r+t)k^2)^2+\Gamma^2}-\frac{i\Gamma}{(\varepsilon-tk^2)^2+\Gamma^2},\\
&&{\cal B}=\frac{\varepsilon-(r+t)k^2}{(\varepsilon-(r+t)k^2)^2+\Gamma^2}-\frac{\varepsilon-tk^2}{(\varepsilon-tk^2)^2+\Gamma^2},\\
&&{\cal C}=\frac{2i\Gamma}{(\varepsilon-(r+t)k^2)^2+\Gamma^2}.
\end{eqnarray}
By taking the limit $\Gamma\rightarrow0$, we find
\begin{eqnarray}
&&{\cal A}\approx 2i\pi\left(\delta(\varepsilon-(r+t)k^2)-\delta(\varepsilon-tk^2)\right),\\
&&{\cal B}\approx0,\\
&&{\cal C}\approx 4i\pi\delta(\varepsilon-(r+t)k^2).
\end{eqnarray}
It is therefore sufficient to compute the commutators applied on the operators $\{\hat X,({\bf k}\cdot{\bf L})^2\}$ and $\hat X$. Let us apply Eq. \eqref{eq:1} to $L_z$. Explicitly, we obtain, 
\begin{eqnarray}
k^2[\hat{\cal H}_k,\hat L_z]^{(2p)}&=&-(rk^2)^{2p}(k_z({\bf k}\cdot\hat {\bf L})-k^2\hat L_z),\\
k^2[\hat{\cal H}_k,\hat L_z]^{(2p+1)}&=&-2i(rk^2)^{2p+1}\left((k_x^2-k_y^2)\hat Q_{xy}+k_zk_x\hat Q_{yz}\right.\nonumber\\&&\left.-k_zky\hat Q_{zx}-2k_xk_y\hat Q_{x^2-y^2}\right),
\end{eqnarray}
and $\left[\hat{\cal H}_k,\{\hat L_z,({\bf k}\cdot\hat{\bf L})^2\}\right]^{(n)}=k^2[\hat{\cal H}_k,\hat L_z]^{(n)}$, $n\in N$.
Since these expressions must be integrated over the spherical Brillouin zone, defining ${\bf k}=k(\cos\varphi\sin\theta,\sin\varphi\sin\theta,\cos\theta)$, we perform the angular averging,

 \begin{eqnarray}
&&\int d\cos\theta d\varphi k^2[\hat{\cal H}_k,\hat L_z]^{(2p)}=\frac{8\pi}{3}k^2(rk^2)^{2p}\hat L_z,\nonumber\\
&&\int d\cos\theta d\varphi k^2[\hat{\cal H}_k,\hat L_z]^{(2p+1)}=0,\nonumber\\
&&\int d\cos\theta d\varphi[\hat{\cal H}_k,\{L_z,({\bf k}\cdot\hat{\bf L})^2\}]^{(2p)}=\frac{8\pi}{3}k^2(rk^2)^{2p}\hat L_z,\nonumber\\
&&\int d\cos\theta d\varphi[\hat{\cal H}_k,\{\hat L_z,({\bf k}\cdot\hat {\bf L})^2\}]^{(2p+1)}=0.\nonumber
\end{eqnarray}
After integration over $\bf k$, we obtain
 \begin{eqnarray}
&&\int\frac{d^3{\bf k}}{(2\pi)^3}\frac{\hat{\cal K}_{L_z}}{\Omega}=\frac{2i\Gamma}{\Omega}\rho_z\hat L_z,\\
&&\int\frac{d^3{\bf k}}{(2\pi)^3}\left[\hat{\cal H}_k,\frac{\hat{\cal K}_{L_z}}{\Omega}\right]^{(n)}=\frac{i\Gamma}{\Omega}\frac{\sqrt{\varepsilon}}{4\pi^2{\cal N}}g(\varepsilon)\rho_z\hat L_z,\\
&&g(\varepsilon)=\frac{(r\varepsilon/t)^{2p}}{t^{3/2}}+\frac{(r\varepsilon/(t+r))^{2p}}{(t+r)^{3/2}}.
\end{eqnarray}
Putting everything together, Eq. \eqref{eq:1} becomes
\begin{eqnarray}\label{eq:a2}
\Omega\rho_z\hat L_z=2i\Gamma\rho_z\hat L_z+i\Gamma\frac{\sqrt{\varepsilon}}{4\pi^2{\cal N}}\sum_{p\geq1}g\left(\frac{\varepsilon}{\Omega}\right)\rho_z\hat L_z,
\end{eqnarray}
We first replace $\Omega=\omega+2i\Gamma\approx 2i\Gamma$, which gives
\begin{eqnarray}\label{eq:a3}
\omega\rho_z\hat L_z&=&i\Gamma\frac{\sqrt{\varepsilon}}{4\pi^2{\cal N}}\sum_{p\geq1}\left(\frac{1}{t^{3/2}}\left[-\left(\frac{\varepsilon}{2\Gamma}\frac{r}{t}\right)^2\right]^p+\frac{1}{(r+t)^{3/2}}\left[-\left(\frac{\varepsilon}{2\Gamma}\frac{r}{r+t}\right)^2\right]^p\right)\rho_z\hat L_z\\
\end{eqnarray}
The summations are of the form $\sum_{n>1} (-x)^n=-\frac{x}{1+x}$, so we obtain
\begin{eqnarray}\label{eq:a4}
\omega\rho_z\hat L_z=-i\frac{\Gamma\sqrt{\varepsilon}}{4\pi^2{\cal N}}\left(\frac{1}{t^{3/2}}\frac{\left(\frac{\varepsilon}{2\Gamma}\frac{r}{t}\right)^2}{1+\left(\frac{\varepsilon}{2\Gamma}\frac{r}{t}\right)^2}+\frac{1}{(t+r)^{3/2}}\frac{\left(\frac{\varepsilon}{2\Gamma}\frac{r}{r+t}\right)^2}{1+\left(\frac{\varepsilon}{2\Gamma}\frac{r}{r+t}\right)^2}\right)\rho_z\hat L_z.
\end{eqnarray}
Finally, we identify the relaxation time, $-i\omega\rho_z=-\rho_z/\tau_s$, and obtain Eq. \eqref{eq:2}. We stress out that, formally, this expression is only valid as long as $\Gamma>\frac{\varepsilon}{2}\frac{r}{t}$, which ensures the convergence of the series. However, to verify the generality of our results, we have also solved Eq. \eqref{eq:rec} directly using standard linear algebra. Such a cumbersome procedure lacks the elegance of the recursive formulation but provides exactly the same result, extending our analytical expression to the full range of momentum relaxation time.
\end{widetext}
\bibliography{Biblio2025}

@article{Szolnoki2017,
   author = {Lénárd Szolnoki and Balázs Dóra and Annamária Kiss and Jaroslav Fabian and Ferenc Simon},
   doi = {10.1103/PhysRevB.96.245123},
   issn = {24699969},
   issue = {24},
   journal = {Physical Review B},
   month = {12},
   pages = {245123},
   publisher = {American Physical Society},
   title = {Intuitive approach to the unified theory of spin relaxation},
   volume = {96},
   year = {2017}
}

@article{Culcer2007,
   author = {Dimitrie Culcer and R. Winkler},
   doi = {10.1103/PhysRevB.76.195204},
   issn = {10980121},
   issue = {19},
   journal = {Physical Review B},
   month = {11},
   pages = {195204},
   title = {Spin polarization decay in spin- 1 2 and spin- 3 2 systems},
   volume = {76},
   year = {2007}
}

@article{Boross2013,
   author = {Péter Boross and Balázs Dóra and Annamária Kiss and Ferenc Simon},
   doi = {10.1038/srep03233},
   issn = {20452322},
   journal = {Scientific Reports},
   pages = {3233},
   title = {A unified theory of spin-relaxation due to spin-orbit coupling in metals and semiconductors},
   volume = {3},
   year = {2013}
}

@article{Aguilar-Pujol2025,
   author = {Montserrat X Aguilar-Pujol and Isabel C Arango and Eoin Dolan and You Ba and Marco Gobbi and Luis E Hueso and Fèlix Casanova},
   journal = {arXiv:2506.06546v2},
   title = {Orbital Hall conductivity and orbital diffusion length of Vanadium thin films by Hanle magnetoresistance},
   year = {2025}
}

@article{Yang2024a,
   author = {Yuhe Yang and Ping Wang and Jiali Chen and Delin Zhang and Chang Pan and Shuai Hu and Ting Wang and Wensi Yue and Cheng Chen and Wei Jiang and Lujun Zhu and Xuepeng Qiu and Yugui Yao and Yue Li and Wenhong Wang and Yong Jiang},
   doi = {10.1038/s41467-024-52824-2},
   issn = {20411723},
   issue = {1},
   journal = {Nature Communications },
   month = {12},
   pages = {8645},
   pmid = {39369005},
   publisher = {Nature Research},
   title = {Orbital torque switching in perpendicularly magnetized materials},
   volume = {15},
   year = {2024}
}

@article{Ding2024,
   author = {Shilei Ding and Min Gu Kang and William Legrand and Pietro Gambardella},
   doi = {10.1103/PhysRevLett.132.236702},
   issn = {10797114},
   issue = {23},
   journal = {Physical Review Letters},
   month = {6},
   pages = {236702},
   pmid = {38905652},
   publisher = {American Physical Society},
   title = {Orbital Torque in Rare-Earth Transition-Metal Ferrimagnets},
   volume = {132},
   year = {2024}
}

@article{Lyalin2023,
   author = {Igor Lyalin and Sanaz Alikhah and Marco Berritta and Peter M. Oppeneer and Roland K. Kawakami},
   doi = {10.1103/PhysRevLett.131.156702},
   issn = {10797114},
   issue = {15},
   journal = {Physical Review Letters},
   month = {10},
   pmid = {37897779},
   publisher = {American Physical Society},
   title = {Magneto-Optical Detection of the Orbital Hall Effect in Chromium},
   volume = {131},
   year = {2023}
}

@article{Huang2025,
   author = {Lin Huang and Da Tian and Liyang Liao and Hongsong Qiu and Hua Bai and Qian Wang and Feng Pan and Caihong Zhang and Biaobing Jin and Cheng Song},
   doi = {10.1002/adma.202402063},
   issn = {15214095},
   issue = {6},
   journal = {Advanced Materials},
   month = {2},
   pages = {2402063},
   pmid = {39707662},
   publisher = {John Wiley and Sons Inc},
   title = {Orbital Current Pumping From Ultrafast Light-driven Antiferromagnetic Insulator},
   volume = {37},
   year = {2025}
}

@article{Hayashi2024,
   author = {Hiroki Hayashi and Dongwook Go and Satoshi Haku and Yuriy Mokrousov and Kazuya Ando},
   doi = {10.1038/s41928-024-01193-1},
   issn = {25201131},
   issue = {8},
   journal = {Nature Electronics},
   month = {8},
   pages = {646-652},
   publisher = {Nature Research},
   title = {Observation of orbital pumping},
   volume = {7},
   year = {2024}
}

@article{Santos2024,
   author = {E. Santos and J. E. Abrão and A. S. Vieira and J. B.S. Mendes and R. L. Rodríguez-Suárez and A. Azevedo},
   doi = {10.1103/PhysRevB.109.014420},
   issn = {24699969},
   issue = {1},
   journal = {Physical Review B},
   month = {1},
   pages = {014420},
   publisher = {American Physical Society},
   title = {Exploring orbital-charge conversion mediated by interfaces with Cu Ox through spin-orbital pumping},
   volume = {109},
   year = {2024}
}

@article{Wang2025,
   author = {Hanchen Wang and Min Gu Kang and Davit Petrosyan and Shilei Ding and Richard Schlitz and Lauren J. Riddiford and William Legrand and Pietro Gambardella},
   doi = {10.1103/PhysRevLett.134.126701},
   issn = {10797114},
   issue = {12},
   journal = {Physical Review Letters},
   month = {3},
   pages = {126701},
   pmid = {40215529},
   publisher = {American Physical Society},
   title = {Orbital Pumping in Ferrimagnetic Insulators},
   volume = {134},
   year = {2025}
}

@article{Sun2025,
   author = {Rui Sun and Yoji Nabei and Aeron McConnell and Xiaotong Zhang and Andrew Comstock and Hana Jones and Rishiram Gyawali and Yuzan Xiong and Ziqi Wang and Jun Liu and Wei Zhang and Dali Sun},
   doi = {10.1063/5.0292745},
   issn = {0021-8979},
   issue = {12},
   journal = {Journal of Applied Physics},
   month = {9},
   pages = {123905},
   title = {Determination of orbital relaxation in Ti/Ni heterostructure via orbital pumping},
   volume = {138},
   url = {https://pubs.aip.org/jap/article/138/12/123905/3364330/Determination-of-orbital-relaxation-in-Ti-Ni},
   year = {2025}
}

@article{Moriya2024,
   author = {Hiroyuki Moriya and Mari Taniguchi and Daegeun Jo and Dongwook Go and Nozomi Soya and Hiroki Hayashi and Yuriy Mokrousov and Hyun Woo Lee and Kazuya Ando},
   doi = {10.1021/acs.nanolett.3c05102},
   issn = {15306992},
   issue = {22},
   journal = {Nano Letters},
   month = {6},
   pages = {6459-6464},
   pmid = {38780051},
   publisher = {American Chemical Society},
   title = {Observation of Long-Range Current-Induced Torque in Ni/Pt Bilayers},
   volume = {24},
   year = {2024}
}

@article{Adamantopoulos2024,
   author = {T. Adamantopoulos and M. Merte and D. Go and F. Freimuth and S. Blügel and Y. Mokrousov},
   doi = {10.1103/PhysRevLett.132.076901},
   issn = {10797114},
   issue = {7},
   journal = {Physical Review Letters},
   month = {2},
   pages = {076901},
   pmid = {38427860},
   publisher = {American Physical Society},
   title = {Orbital Rashba Effect as a Platform for Robust Orbital Photocurrents},
   volume = {132},
   year = {2024}
}

@article{Bose2023,
   author = {Arnab Bose and Fabian Kammerbauer and Rahul Gupta and Dongwook Go and Yuriy Mokrousov and Gerhard Jakob and Mathias Kläui},
   doi = {10.1103/PhysRevB.107.134423},
   issn = {24699969},
   issue = {13},
   journal = {Physical Review B},
   month = {4},
   pages = {134423},
   publisher = {American Physical Society},
   title = {Detection of long-range orbital-Hall torques},
   volume = {107},
   year = {2023}
}

@article{Veneri2024,
   author = {Alessandro Veneri and Tatiana G. Rappoport and Aires Ferreira},
   doi = {10.1103/PhysRevLett.134.136201},
   issn = {10797114},
   issue = {13},
   journal = {Physical Review Letters},
   month = {4},
   pages = {136201},
   pmid = {40250347},
   publisher = {American Physical Society},
   title = {Extrinsic Orbital Hall Effect: Orbital Skew Scattering and Crossover between Diffusive and Intrinsic Orbital Transport},
   volume = {134},
   year = {2025}
}

@article{Gao2025,
   author = {Weiguang Gao and Liyang Liao and Hironari Isshiki and Nico Budai and Junyeon Kim and Hyun Woo Lee and Kyung Jin Lee and Dongwook Go and Yuriy Mokrousov and Shinji Miwa and Yoshichika Otani},
   doi = {10.1038/s41467-025-61602-7},
   issn = {20411723},
   issue = {1},
   journal = {Nature Communications },
   month = {12},
   pages = {6380},
   pmid = {40640152},
   publisher = {Nature Research},
   title = {Nonlocal electrical detection of reciprocal orbital Edelstein effect},
   volume = {16},
   year = {2025}
}

@article{Ledesma-Martin2025,
   author = {José Omar Ledesma-Martin and Edgar Galindez-Ruales and Sachin Krishnia and Felix Fuhrmann and Minh Duc Tran and Rahul Gupta and Marcel Gasser and Dongwook Go and Akashdeep Kamra and Gerhard Jakob and Yuriy Mokrousov and Mathias Kläui},
   doi = {10.1021/acs.nanolett.4c06056},
   issn = {15306992},
   journal = {Nano Letters},
   month = {2},
   publisher = {American Chemical Society},
   title = {Nonreciprocity in Magnon Mediated Charge-Spin-Orbital Current Interconversion},
   year = {2025}
}

@article{Ning2025,
   author = {Xiaobai Ning and A. Pezo and Kyoung-Whan Kim and Weisheng Zhao and Kyung-Jin Lee and Aurélien Manchon},
   doi = {10.1103/PhysRevLett.134.026303},
   issn = {0031-9007},
   issue = {2},
   journal = {Physical Review Letters},
   month = {1},
   pages = {026303},
   title = {Orbital Diffusion, Polarization, and Swapping in Centrosymmetric Metals},
   volume = {134},
   url = {https://link.aps.org/doi/10.1103/PhysRevLett.134.026303},
   year = {2025}
}

@article{Gupta2025,
   author = {Rahul Gupta and Chloé Bouard and Fabian Kammerbauer and J. Omar Ledesma-Martin and Arnab Bose and Iryna Kononenko and Sylvain Martin and Perrine Usé and Gerhard Jakob and Marc Drouard and Mathias Kläui},
   doi = {10.1038/s41467-024-55437-x},
   issn = {20411723},
   issue = {1},
   journal = {Nature Communications },
   month = {12},
   pages = {130},
   publisher = {Nature Research},
   title = {Harnessing orbital Hall effect in spin-orbit torque MRAM},
   volume = {16},
   year = {2025}
}

@article{Sala2023,
   author = {Giacomo Sala and Hanchen Wang and William Legrand and Pietro Gambardella},
   doi = {10.1103/PhysRevLett.131.156703},
   issn = {10797114},
   issue = {15},
   journal = {Physical Review Letters},
   month = {10},
   pages = {156703},
   pmid = {37897743},
   publisher = {American Physical Society},
   title = {Orbital Hanle Magnetoresistance in a 3d Transition Metal},
   volume = {131},
   year = {2023}
}

@article{Sohn2024,
   author = {Jeonghun Sohn and Jongjun M. Lee and Hyun Woo Lee},
   doi = {10.1103/PhysRevLett.132.246301},
   issn = {10797114},
   issue = {24},
   journal = {Physical Review Letters},
   month = {6},
   pages = {246301},
   pmid = {38949365},
   publisher = {American Physical Society},
   title = {Dyakonov-Perel-like Orbital and Spin Relaxations in Centrosymmetric Systems},
   volume = {132},
   year = {2024}
}

@article{Choi2023,
   author = {Young Gwan Choi and Daegeun Jo and Kyung Hun Ko and Dongwook Go and Kyung Han Kim and Hee Gyum Park and Changyoung Kim and Byoung Chul Min and Gyung Min Choi and Hyun Woo Lee},
   doi = {10.1038/s41586-023-06101-9},
   issn = {14764687},
   issue = {7968},
   journal = {Nature},
   month = {7},
   pages = {52-56},
   pmid = {37407680},
   publisher = {Nature Research},
   title = {Observation of the orbital Hall effect in a light metal Ti},
   volume = {619},
   year = {2023}
}

@article{Han2022,
   author = {Seungyun Han and Hyun Woo Lee and Kyoung Whan Kim},
   doi = {10.1103/PhysRevLett.128.176601},
   issn = {10797114},
   issue = {17},
   journal = {Physical Review Letters},
   month = {4},
   pmid = {35570433},
   publisher = {American Physical Society},
   title = {Orbital Dynamics in Centrosymmetric Systems},
   volume = {128},
   year = {2022}
}

@article{Go2023a,
   author = {Dongwook Go and Daegeun Jo and Kyoung-Whan Kim and Soogil Lee and Min-Gu Kang and Byong-Guk Park and Stefan Blügel and Hyun-Woo Lee and Yuriy Mokrousov},
   doi = {10.1103/physrevlett.130.246701},
   issn = {0031-9007},
   issue = {24},
   journal = {Physical Review Letters},
   month = {6},
   publisher = {American Physical Society (APS)},
   title = {Long-Range Orbital Torque by Momentum-Space Hotspots},
   volume = {130},
   year = {2023}
}

@article{Hayashi2023,
   author = {Hiroki Hayashi and Daegeun Jo and Dongwook Go and Tenghua Gao and Satoshi Haku and Yuriy Mokrousov and Hyun Woo Lee and Kazuya Ando},
   doi = {10.1038/s42005-023-01139-7},
   issn = {23993650},
   issue = {1},
   journal = {Communications Physics},
   month = {12},
   publisher = {Nature Research},
   title = {Observation of long-range orbital transport and giant orbital torque},
   volume = {6},
   year = {2023}
}

@article{Salemi2021,
   author = {Leandro Salemi and Marco Berritta and Peter M. Oppeneer},
   doi = {10.1103/PhysRevMaterials.5.074407},
   issn = {24759953},
   issue = {7},
   journal = {Physical Review Materials},
   month = {7},
   publisher = {American Physical Society},
   title = {Quantitative comparison of electrically induced spin and orbital polarizations in heavy-metal/ 3d -metal bilayers},
   volume = {5},
   year = {2021}
}

@article{Salemi2022,
   author = {Leandro Salemi and Peter M. Oppeneer},
   doi = {10.1103/physrevmaterials.6.095001},
   journal = {Physical Review Materials},
   pages = {095001},
   publisher = {American Physical Society},
   title = {First-principles theory of intrinsic spin and orbital Hall and Nernst effects in metallic monoatomic crystals},
   volume = {6},
   year = {2022}
}

@article{Pezo2022,
   author = {Armando Pezo and Diego Garcia Ovalle and Aurelien Manchon},
   doi = {10.1103/PhysRevB.106.104414},
   issn = {24699969},
   journal = {Physical Review B},
   pages = {104414},
   publisher = {American Physical Society},
   title = {Orbital Hall effect in crystals: inter-atomic versus intra-atomic contributions},
   volume = {106},
   url = {http://arxiv.org/abs/2201.05807},
   year = {2022}
}

@article{Ding2022,
   author = {Shilei Ding and Zhongyu Liang and Dongwook Go and Chao Yun and Mingzhu Xue and Zhou Liu and Sven Becker and Wenyun Yang and Honglin Du and Changsheng Wang and Yingchang Yang and Gerhard Jakob and Mathias Kläui and Yuriy Mokrousov and Jinbo Yang},
   doi = {10.1103/physrevlett.128.067201},
   issn = {0031-9007},
   journal = {Physical Review Letters},
   pages = {067201},
   pmid = {35213174},
   publisher = {American Physical Society},
   title = {Observation of the Orbital Rashba-Edelstein Magnetoresistance},
   volume = {128},
   url = {https://doi.org/10.1103/PhysRevLett.128.067201},
   year = {2022}
}

@article{Lee2021b,
   author = {Dongjoon Lee and Dongwook Go and Hyeon Jong Park and Wonmin Jeong and Hye Won Ko and Deokhyun Yun and Daegeun Jo and Soogil Lee and Gyungchoon Go and Jung Hyun Oh and Kab Jin Kim and Byong Guk Park and Byoung Chul Min and Hyun Cheol Koo and Hyun Woo Lee and Ouk Jae Lee and Kyung Jin Lee},
   doi = {10.1038/s41467-021-26650-9},
   issn = {20411723},
   journal = {Nature Communications},
   pages = {6710},
   pmid = {34795204},
   publisher = {Springer US},
   title = {Orbital torque in magnetic bilayers},
   volume = {12},
   year = {2021}
}

@article{Lee2021a,
   author = {Soogil Lee and Min Gu Kang and Dongwook Go and Dohyoung Kim and Jun Ho Kang and Taekhyeon Lee and Geun Hee Lee and Jaimin Kang and Nyun Jong Lee and Yuriy Mokrousov and Sanghoon Kim and Kab Jin Kim and Kyung Jin Lee and Byong Guk Park},
   doi = {10.1038/s42005-021-00737-7},
   isbn = {4200502100},
   issn = {23993650},
   journal = {Communications Physics},
   pages = {234},
   publisher = {Springer US},
   title = {Efficient conversion of orbital Hall current to spin current for spin-orbit torque switching},
   volume = {4},
   year = {2021}
}

@article{Canonico2020,
   author = {Luis M. Canonico and Tarik P. Cysne and Alejandro Molina-Sanchez and R. B. Muniz and Tatiana G. Rappoport},
   doi = {10.1103/PhysRevB.101.161409},
   issn = {24699969},
   journal = {Physical Review B},
   pages = {161409(R)},
   title = {Orbital Hall insulating phase in transition metal dichalcogenide monolayers},
   volume = {101},
   year = {2020}
}

@article{Bhowal2021,
   author = {Sayantika Bhowal and Giovanni Vignale},
   doi = {10.1103/PhysRevB.103.195309},
   issn = {24699969},
   journal = {Physical Review B},
   pages = {195309},
   title = {Orbital Hall effect as an alternative to valley Hall effect in gapped graphene},
   volume = {103},
   year = {2021}
}

@article{Go2021,
   author = {Dongwook Go and Daegeun Jo and Tenghua Gao and Kazuya Ando and Stefan Blügel and Hyun Woo Lee and Yuriy Mokrousov},
   doi = {10.1103/physrevb.103.l121113},
   issn = {23318422},
   journal = {Physical Review B},
   pages = {L121113},
   title = {Orbital Rashba effect in surface oxidized Cu film},
   volume = {103},
   year = {2021}
}

@article{Ding2020,
   author = {Shilei Ding and Andrew Ross and Dongwook Go and Lorenzo Baldrati and Zengyao Ren and Frank Freimuth and Sven Becker and Fabian Kammerbauer and Jinbo Yang and Gerhard Jakob and Yuriy Mokrousov and Mathias Kläui},
   doi = {10.1103/PhysRevLett.125.177201},
   issn = {10797114},
   journal = {Physical Review Letters},
   pages = {177201},
   pmid = {33156648},
   publisher = {American Physical Society},
   title = {Harnessing Orbital-to-Spin Conversion of Interfacial Orbital Currents for Efficient Spin-Orbit Torques},
   volume = {125},
   url = {https://doi.org/10.1103/PhysRevLett.125.177201},
   year = {2020}
}

@article{Manchon2020,
   author = {Guilhem Manchon and Sumit Ghosh and Cyrille Barreteau and Aurélien Manchon},
   doi = {10.1103/PhysRevB.101.174423},
   journal = {Physical Review B},
   pages = {174423},
   publisher = {American Physical Society},
   title = {Semirealistic tight-binding model for spin-orbit torques},
   volume = {101},
   year = {2020}
}

@article{Bernevig2005b,
   author = {B Andrei Bernevig and Taylor L Hughes and Shou-cheng Zhang},
   doi = {10.1103/PhysRevLett.95.066601},
   journal = {Physical Review Letters},
   pages = {066601},
   title = {Orbitronics : The Intrinsic Orbital Current in p -Doped Silicon},
   volume = {95},
   year = {2005}
}

@article{Jo2018,
   author = {Daegeun Jo and Dongwook Go and Hyun-woo Lee},
   doi = {10.1103/PhysRevB.98.214405},
   journal = {Physical Review B},
   pages = {214405},
   publisher = {American Physical Society},
   title = {Gigantic intrinsic orbital Hall effects in weakly spin-orbit coupled metals},
   volume = {98},
   year = {2018}
}

@article{Go2017,
   author = {Dongwook Go and Jan-philipp Hanke and Patrick M Buhl and Frank Freimuth and Gustav Bihlmayer and Hyun-woo Lee and Yuriy Mokrousov and Stefan Blügel},
   doi = {10.1038/srep46742},
   journal = {Scientific Reports},
   pages = {46742},
   publisher = {Nature Publishing Group},
   title = {Toward surface orbitronics : giant orbital magnetism from the orbital Rashba effect at the surface of sp-metals},
   volume = {7},
   url = {http://dx.doi.org/10.1038/srep46742},
   year = {2017}
}

@article{Go2020b,
   author = {Dongwook Go and Frank Freimuth and Jan-philipp Hanke and Fei Xue and Olena Gomonay and Kyung-jin Lee and Stefan Blügel and Paul M Haney and Hyun-woo Lee and Yuriy Mokrousov},
   doi = {10.1103/PhysRevResearch.2.033401},
   issn = {0031-899X},
   journal = {Physical Review Research},
   pages = {33401},
   publisher = {American Physical Society},
   title = {Theory of current-induced angular momentum transfer dynamics in spin-orbit coupled systems},
   volume = {2},
   url = {https://doi.org/10.1103/PhysRevResearch.2.033401},
   year = {2020}
}

@article{Go2018,
   author = {Dongwook Go and Daegeun Jo and Changyoung Kim and Hyun Woo Lee},
   doi = {10.1103/PhysRevLett.121.086602},
   isbn = {6082744330},
   issn = {10797114},
   journal = {Physical Review Letters},
   pages = {086602},
   publisher = {American Physical Society},
   title = {Intrinsic Spin and Orbital Hall Effects from Orbital Texture},
   volume = {121},
   url = {https://doi.org/10.1103/PhysRevLett.121.086602},
   year = {2018}
}

@article{Elliott1954,
   author = {R. J. Elliott},
   journal = {Phys. Rev.},
   pages = {266},
   title = {Theory of the Effect of Spin-Orbit Coupling on Magnetic Resonance in Some Semiconductors},
   volume = {96},
   year = {1954}
}

@inbook{Yafet1963,
   author = {Y. Yafet},
   editor = {F. Seitz and D. Turnbull},
   booktitle = {Solid State Physics, Vol. 14},
   pages = {2},
   publisher = {Academic, New York},
   title = {g factors and spin-lattice relaxation of conduction electrons},
   year = {1963}
}

@article{Rammer1986,
   author = {S Rammer and H. Smith},
   journal = {Reviews of Modern Physics},
   pages = {323},
   title = {Quantum field theoretical methods in transport theory of metals},
   volume = {58},
   year = {1986}
}

@article{Dyakonov1972,
   author = {MI Dyakonov and VI Perel},
   journal = {Soviet Physics Solid State},
   pages = {3023},
   title = {Spin relaxation of conduction electrons in noncentrosymmetric semiconductors},
   volume = {13},
   year = {1972}
}

@book{Dyakonov2008a,
   city = {Berlin},
   editor = {MI Dyakonov},
   isbn = {9783540788195},
   journal = {Series in Solid-State Sciences},
   publisher = {Springer-Verlag Berlin},
   title = {Spin physics in semiconductors},
   url = {http://books.google.com/books?hl=en&lr=&id=PJ91JeCvcGoC&oi=fnd&pg=PA441&dq=Springer+Series+in&ots=nencEFTtjl&sig=u1ApyrbzpCtTEE-bnKQpQdqE-dk http://link.springer.com/content/pdf/10.1007/978-3-540-78820-1.pdf},
   year = {2008}
}

@article{Fabian2007,
   author = {J. Fabian and A. Matos-Abiague and C. Ertler and P. Stano and I. Zutic},
   issue = {4},
   journal = {acta physica slovaca},
   pages = {565-907},
   title = {Semiconductor Spintronics},
   volume = {57},
   year = {2007}
}

@article{Wang2014,
   author = {Xuhui Wang and Christian Ortiz Pauyac and Aurélien Manchon},
   journal = {Physical Review B},
   pages = {054405},
   title = {Spin-orbit-coupled transport and spin torque in a ferromagnetic heterostructure},
   volume = {89},
   url = {http://link.aps.org/doi/10.1103/PhysRevB.89.054405},
   year = {2014}
}

\end{document}